# 3D residual stress field in arteries: novel inverse method based on optical full-field measurements

Badel Pierre (a)*, Genovese Katia (b), Avril Stéphane (a)

(a) Center for Health Engineering, Ecole des Mines de Saint Etienne, LCG-CNRS UMR5146, 158 cours Fauriel, 42023 Saint Etienne, France

(b) Dept. of Engineering and Environmental Physics, Università degli Studi della Basilicata, Viale dell'Ateneo Lucano 10, 85100 Potenza, Italy

* Corresponding author:
Tel: +33(0)477420260
Fax: +33(0)477499755
badel@emse.fr
http://www.cis.emse.fr/

ABSTRACT

Arterial tissue consists of multiple structurally important constituents which have individual material properties and associated stress-free configurations that evolve over time. This gives rise to residual stresses contributing to the homeostatic state of stress *in vivo* as well as adaptations to perturbed loads, disease, or injury. The existence of residual stresses in an intact but load-free excised arterial segment suggests compressive and tensile stresses respectively in the inner outer walls. Accordingly, an artery ring springs open into a sector after a radial cut. The measurement of the opening angle is commonly used to deduce the residual stresses, which are the stresses required to close back the ring.

The opening angle method provides an average estimate of circumferential residual stresses but it gives no information on local distributions through the thickness and along the axial direction. To address this lack, a new method is proposed in this paper to derive maps of residual stresses using an approach based on the contour method. A piece of freshly excised tissue is carefully cut into the specimen and the local distribution of residual strains and stresses is determined from whole-body Digital Image Correlation measurements using an inverse approach based on a finite element model.

1.      **Introduction**

It has been evidenced for several decades that even removed from the body, therefore load-free, the arterial wall is not stress-free. An unloaded ring-like segment of an artery springs open to form an arc when it is cut radially [1-3]. Arterial tissue is similar to a laminated fibre-reinforced composite material consisting in multiple types of structurally important constituents, each of which may have individual material properties [4] and associated stress-free configurations that may evolve over time. This gives rise to residual stresses which are thought to contribute to the homeostatic state of stress *in vivo* as well as adaptations to perturbed loads, diseases, or injury [1,2,5]. Characterizing residual stresses in blood vessel is an important challenge towards the understanding of the vascular function and associated diseases. It is also a key feature for properly defining theoretical or numerical mechanical problems used to study healthy or diseased arteries. In such problems the knowledge of the stress-free configuration is fundamental. Furthermore, the development of tissue engineering requires a better understanding of such adaptation phenomena, in order to reproduce them.

The origin and mechanisms giving rise to residual stresses are still at the stage of hypotheses. For instance, Cardamone et. al. [6] suggest that axial pre-stretch and residual stresses arise in arteries largely due to the deposition of stable, highly elastic, elastin during vascular development and the continual turnover of collagen and smooth muscle at consistent, preferred stretches within an important ground substance matrix. Over time, elastin will tend to stretch, while collagen is continuously re-synthesized at the *in vivo* length, leading to different stress states when the artery is unloaded. Azeloglu et. al. [7] investigated, both numerically and experimentally, the regulating role of proteoglycans present in the arterial wall. Due to the inhomogeneous distribution of these negatively charged proteoglycans enmeshed within the collagen matrix, the resulting Donnan osmotic pressure would vary through the wall thickness. This would lead to an inhomogeneous swelling stress field in the solid matrix, which significantly affects the opening angle observed experimentally. In addition, the residual state of arteries has been shown to be location-dependent and be influenced by an amount of other factors, a review of which can be found in [5].

The role of residual strains and stresses is still not clear. Several authors have shown that although residual stresses are small compared with *in vivo* wall stresses they have a strong influence on the *in vivo* stress distribution. In particular, they avoid the stress concentration that would occur at the innermost layer of the arterial wall [3]. The residual stresses seem to have the effect of homogenizing the circumferential stress within each layer in the physiological load state [2,3]. Hence, they would play a major role in maintaining a so-called homeostatic stress state. This homeostatic state being perturbed during the development of vasculature or during the apparition of diseases like hypertension or wall defects, the evolution of residual stresses is likely to play a major role in remodeling and growth processes. Understanding the role of residual stresses is crucial in consideration of such fundamental issues.

Previous experimental work regarding residual stresses has been mainly devoted to the opening angle method, which is a fairly straightforward method (see the review [5] and references therein). Indeed, the existence of residual stress is manifest in the springing open of an arterial ring when it is cut in the radial direction. The measurement of the opening angle is used to deduce the residual stresses, which are actually the stresses required to close back the ring. In the majority of experiments, rings from excised vessels are placed in a neutrally buoyant medium and an image of the unloaded state is captured. The ring is then cut, replaced in the medium and an image of the putative zero-stress state is captured. Required geometrical parameters are determined by manual or computerized measurement of the two images. Though the cut specimens rarely take the form of circular segments, this technique remains popular, presumably because of its technical simplicity.

The ring-opening of the arterial segment was long considered to be a plane strain problem [8], the axial stretch being assumed equal to one. However it was shown that residual stresses are also present in the axial direction. Recently, Holzapfel et. al. [9] quantified the non-zero axial residual strain on longitudinal strips using an experimental protocol similar to that used for transversal sections. Therefore the study of residual stresses has to be considered as a 3D problem [9,10].

In addition to this difficulty, note that the opening angle method provides an average estimate of the circumferential residual stresses but it gives no information on either the local distribution through the thickness or along the axial direction. Moreover, the procedure relies on the assumption that the residual stresses in the ring are completely released after one single radial cut. This assumption is questionable as each constituent of the wall has its own constitutive properties and *in vivo* strain state, which will likely leave different stress states when the segment is unloaded. For this reason, the opening angle method is still to be improved. Moreover, Zhang et al. [11] showed artifacts that may be induced by viscoelasticity. Recently, Ohayon et al. [12] also used the opening angle method for an artery with an atheromatous plaque. This yielded an interesting study about the definition of plaque vulnerability. However, the approach is somehow questionable as the heterogeneity and localization of residual stresses in a plaque may not be fully revealed by the method. Actually, the opening angle method is well suited when the distribution of residual stresses is axisymmetric. There is still a lack of a more local method that may be used for identifying the residual stresses in an artery of any shape, and especially with a plaque.

Several authors have tried to address this lack. Considering the alternation of elastin rich layers and smooth muscle rich layers, Matsumoto et. al. [4] showed that such inhomogeneity of the arterial wall leads to strong heterogeneities in the residual stress field. Han and Fung [13] used markers to track local radial and circumferential deformations during the opening angle experiment. Holzapfel et al. [9] considered the existence of layer-specific residual stretch by separating mechanically the different layers after performing the opening angle experiment, and thus obtained layer-specific values of opening angle, proving that the assumption of homogeneity is insufficient to study the residual state of the whole arterial wall.

In order to gain a new insight into this important question, we propose a methodology to derive the 3D local distribution of the residual stresses. It uses an approach based on the most recent methods developed for the measurement of residual stresses in laminated fibre-reinforced composite materials.

Indeed, reconstruction of residual stresses has always been an important topic of research regarding engineering applications. Many techniques have been developed by material engineers to measure residual stresses. For laminated composite materials for instance, at a ply (or macroscopic) scale, the hole-drilling method [14], the layer removal method and the compliance method may be used [15]. The compliance method applied to filament wound tubes consists in sectioning the tubes and measuring the change in strains on the external and internal surfaces with two biaxial strain gages, one bonded to each side of the tube wall. This leads to the estimation of the axial and circumferential internal bending moments of filament wound tubes. Using the hole-drilling or other removal approaches in combination with full displacement field optical techniques may provide some advantages with respect to the traditional use of strain gauges, in particular higher sensitivity and no-contact measurement [16].

Removal methods combined to the measurement of local deformations are referred to as the contour method [17,18]. We propose here its application to arterial tissue, with the particularity that deformation measurement focuses on the removed part of the material. The approach combines whole-body Digital Image Correlation (DIC) measurements and the resolution of an inverse problem through a finite element model method. Small pieces of freshly excised tissue are carefully cut out of the specimen and the local 3D distribution of residual stresses is inversely determined from both the study of their deformation after removal and finite element (FE) models of the whole removal procedure.

## 2. Materials and methods

Through this section, the reader can refer to the outline of the method provided in Fig. 1. Shortly, it consists of 3 main steps:

(i) cut out a cube-shaped part of a bovine aorta nailed with nine fine needles to a foam substrate,

(ii) measure the deformation induced by needle removal and deduce the stress field in the tissue just after cutting the cube shape,

(iii) relate this stress field to the initial residual stresses of the artery through a FE model and calibrate this FE model for identifying the initial residual stresses of the artery.

2.1. Materials

A fairly straight segment of a fresh bovine descending aorta (about 40 mm length, 30 mm outer diameter, 4 mm thickness) was obtained from a local slaughterhouse. Loose connective tissues were carefully removed in physiological solution and the artery was put vertically on a plastic cap (Fig. 2a). A ring of the artery was used to calculate the circumferential residual strains with the opening angle method. A polyurethane expandable sealant foam was sprayed into the artery lumen while keeping it moist by means of cotton pads soaked in physiological solution. During the curing time (about 10 min), the foam gently expands and fits closely the artery inner geometry penetrating the branches. A preliminary test carried out by filling a latex glove finger with the foam provided no evidence of stress generation during its expansion and hardening. Once the foam hardened, eight acupuncture needles (0.25 mm diameter, 20 mm length) were used to delimit square areas on the arterial external surface. The needles penetrated through the arterial wall thickness and through the foam (for about 15 mm) allowing thus to maintain the selected pieces in a configuration where they still undergo a fraction of the residual stresses. A scalpel was then used to cut out the specimen (Fig. 2b), first by removing the remaining part of the artery (note how the artery opened up in response to the radial cuts) and finally by cutting the foam (Fig. 2c). It is important to note that after this first step, only a fraction of the initial residual stresses remains in the specimen, which is in itself a strong mark of this initial residual stress field. This important aspect will be taken into account and post-corrected in the final numerical reconstruction. Two additional needles were then used to handle the specimen during all the stages of the experimental protocol. The samples were checked carefully in order to verify the adhesion of the inner arterial surface to the foam and then were sprayed with black Indian ink by using a fine tipped airbrush. This allowed creating the speckle random pattern needed for the subsequent DIC-based measurement [19].

2.2. Experimental set-up and protocol for releasing stresses

In order to obtain the displacement fields induced by the needles removal, the aim of the experimental test was to retrieve the shape of the sample before and after removing the needles. A DIC-based whole-body measurement was obtained by positioning the sample on the top of a ½" steel post fixed onto a rotational stage (Fig. 3). A single camera was used to capture a 1 min video sequence of the sample throughout a full 360° rotation. In this way, each couple of frames of the video sequence (of about 200 frames) can be considered two views of a lateral DIC stereo-system and can be thus processed on the basis of the classic Stereo-Photogrammetry principles [20]. In particular, calibration was performed with the modified DLT method [21] that allows retrieving the stereo-system parameters by using a single image. To this goal, a square dot pattern was glued around the steel post allowing the presence of the calibration pattern in all the images of the sequence. Moreover, in this way, it is made possible to automatically reconstruct and merge the point clouds into a single global reference system. All the codes needed for calibration, DIC-matching, reconstruction and merging were compiled into the Matlab® environment.

2.3. Deformation measurement

A first video sequence of the sample was recorded to obtain its shape in the needle-maintained configuration (i.e. with a remaining fraction of the initial residual stresses, Fig. 4a). Then, without moving the sample, the needles were carefully removed with tweezers and the central needle was used to slightly lift the sample away from the foam, thereby ensuring no contact between the foam and the inner surface of the artery. Once this supposedly stress-free configuration was obtained (Fig. 4b), a second video sequence was recorded.

Due to the large amount of frames recorded, it was possible to choose two closely corresponding views of the needle-maintained and stress-free configurations to match them via DIC. Image processing was performed by considering a measurement grid of about 2000 evenly spaced points for each face and by using template and analysis subset sizes of $21\times 21$ pixel$^2$ and $41\times 41$ pixel$^2$, respectively. Only point pairs matched with a correlation coefficient above 0.98 were considered for subsequent analysis. Eight couples of stereo-systems were considered for reconstructing the sample lateral surface all over 360° (four for the edges and four for the faces). The accuracy of the data overlap between the contiguous reconstructed surfaces (of the order of $10^{-2}$ mm) made it not necessary to consider a larger number of stereo-views.

Figure 5a shows the cloud of data points obtained for the stress-free configuration. In this case, it was also possible to reconstruct the top and the bottom of the arterial segment. Being load-free, these two surfaces do not need to be contoured in the needle-maintained configuration. Clouds of data points so obtained were used as input to build the finite-element (FE) model.

2.4. Numerical reconstruction of the residual stress field

**Overview of the method**

The approach proposed to estimate the initial residual stress field (denoted $\underline{\sigma}^0$) in the artery is based on an inverse method. A FE model of the experimental procedure is set up to compute the deformation between the initial configuration (with the residual stresses $\underline{\sigma}^0$) and the needle-maintained configuration (with still a fraction of the residual stresses remaining thanks to the needles, denoted as the remaining stresses $\underline{\sigma}^1$). An output of this model is the remaining residual stress field $\underline{\sigma}^{1\,mod}$ in the needle-maintained configuration, i.e. after cutting out the sample, and an input of the model is the unknown initial residual stress field $\underline{\sigma}^0$.

Such output $\underline{\sigma}^1$ may be also derived from experimental measurements. Indeed, by prescribing the measured displacements as boundary conditions of a second FE model, an experimentally driven estimate of this intermediary stress field, denoted $\underline{\sigma}^{1\,exp}$, can be deduced. The procedure for obtaining $\underline{\sigma}^{1\,exp}$ is detailed further.

Therefore, the objective is to retrieve the initial residual stress field $\underline{\sigma}^0$ through an inverse approach. This is achieved iteratively until a good match is reached between $\underline{\sigma}^{1\,mod}$ and $\underline{\sigma}^{1\,exp}$.

**Numerical reconstruction of the needle-maintained stress field**

Based on the shape and displacement experimental measurements described in section 2.3, a FE model was built in order to obtain the three-dimensional residual strain field as well as an estimation of the residual stress field $\underline{\sigma}^{1\,exp}$ in the needle-maintained configuration. The principle was to create a model of the stress-free specimen and to apply the opposite of the displacement field which was measured experimentally. Thus, the initial state of the FE model corresponded to the final state of the experimental procedure, i.e. the stress-free state, whereas the deformed state of the FE model corresponded to the pre-measurement experimental state, i.e. the needle-maintained state of the sample.

In order to avoid noise or defect issues in the data, to have an efficient computational model and improve the post-analysis, the initial geometry of the FE model had to be idealized. Each face of the cube-like geometry of the specimen was considered as planar and remaining planar during the test. The plane modeling each face analyzed by DIC was derived by least-squares regression from the coordinates of the points obtained by the DIC analysis. The two remaining faces (i.e. the internal and external surfaces of the arterial wall) were approximated by planes obtained from the shape measurement because no DIC analysis could be performed on the points of these two faces. The initial geometry thus obtained is shown in Fig. 5b. The mesh was made of 8000 elements and 35853 degrees of freedom.

To determine the boundary conditions to be applied to the FE model, the same treatment was performed on the deformed data obtained from the DIC analysis. For each face, the plane was derived by least-squares regression from the cloud of deformed points following the same procedure as for the undeformed specimen. From the knowledge of the undeformed and deformed planes a rigid body motion of the plane could be determined. Note that the motion components that remain undetermined – i.e. the rotation about the normal direction of the plane and the translations along its tangent directions – were assumed to be zero. Then, the boundary conditions were derived from the motion of the planes. The condition applied to the FE nodes of each face only consisted of a kinematic condition imposing that the nodes would remain on the plane corresponding to this face. Such boundary conditions may cause excessive deformation of the specimen along the edges and corners. Therefore, to avoid stress concentrations at these sites, the nodes belonging to the edges were left free. Nevertheless, due to these issues, it is expected that the mechanical fields computed close to the corners, i.e. close to the internal and external surfaces of the arterial wall, may be erroneous.

Note that it was possible to directly use the raw geometry and displacement boundary conditions obtained for the experimental procedure. This approach was performed as preliminary work [22] but it presents severe FE convergence issues, likely due to local geometry and displacement heterogeneities. In addition the results seemed to be strongly affected by local heterogeneities and noise in the data.

It is necessary to provide a constitutive equation for the FE calculation to be performed and then the whole 3D strain and stress fields to be extracted. The constitutive model of the arterial material being unknown *a priori*, it was chosen to use a simple constitutive model. Indeed, the analysis is performed between the stress-free and the residually-stressed states, which means that the orders of magnitude of stresses, as well as that of strains, are very low with respect to the physiological conditions. For this reason, assuming the behavior is linear or quasi-linear in such conditions seems to be a valid assumption. Regarding anisotropy, it is neglected as a first approximation, assuming that stiffness is very low and does not vary much with orientation at such strains. Note, however, that this approach allows to refine and complexify the model as desired when the required information is available.

A neo-Hookean model was therefore implemented for its simplicity. This hyper-elastic constitutive model is governed by a strain energy density function *W* of the following form:

$$W = \frac{G}{2}\left(\overline{I_1} - 3\right) + \frac{K}{2}(J-1)^2 \tag{1}$$

where $\overline{I_1}$ is the first deviatoric strain invariant, $J = \text{tr}\left(\underline{\underline{F}}.\underline{\underline{F}}^T\right)$ is the volume ratio computed from the gradient tensor $\underline{\underline{F}}$. The behavior of the material is driven by the parameters $G$ and $K$. For infinitesimal deformations, $K$ denotes the bulk modulus (or compressibility modulus) and $G$ denotes the shear modulus. The values of $G$ and $K$ being unknown, they were set according to the range of values reported in the literature for human and mammals aortas [23-27,10], also assuming the material to be quasi incompressible [28,29]. Most of the mentioned studies provide data in terms of incremental Young's modulus and/or other constitutive models, therefore the values of $G$ and $K$ were obtained by linearizing and identifying the parameters when needed. Quasi-incompressiblity was assumed to correspond to an equivalent Poisson's ratio of 0.49. The following values were used: $G$ = 8 kPa and $K$ = 800 kPa. Tests performed on similar arteries at strain of less than 5% have confirmed these values.

The problem was solved using an implicit scheme.

The strain and stress fields were then reconstructed through FE resolution, these fields correspond to the needle-maintained configuration. In order to evaluate their distribution through the specimen and make the results easier to interpret, average values were computed within ten layers of the specimen, corresponding to circumferential layers of the arterial wall. To this aim, the axial and circumferential components were averaged over each group $i$ of elements with coordinates comprised between r = $r_i$ and r = $r_{i+1}$, spanning from the innermost radius to the outermost radius of the specimen. This stress distribution, function of r, will be used as the experimental needle-maintained stress field in the following and denoted $\underline{\underline{\sigma}}^{1\,\text{exp}}$.

**Inverse evaluation of the residual stress field**

The previous step of the method, detailed above, was performed to obtain the stress field $\underline{\underline{\sigma}}^{1\,\text{exp}}$ in the needle-maintained configuration, as a function of the radial component r. This stress field was used in the following to obtain the initial residual stress field $\underline{\underline{\sigma}}^0$ in the artery via an inverse method. This method is based on three main tools: a FE model of the cutting-out step of the experiment, a $\underline{\underline{\sigma}}^1$-based cost function to be minimized and an iterative updating of the initial residual stress field (see Fig. 6).

The FE model presented in the previous section was further developed with eight needles inserted through the continuum according to the same pattern as the experiment (see Fig. 6b). These needles were modeled as 0.25 mm diameter beams with an elastic constitutive model characterized by a Young's modulus of 210 GPa and a Poisson's ratio of 0.3. The beams were assumed to be perfectly tied to the arterial material, and to be fixed at the inner surface of the arterial wall. The initial conditions of this model consisted of the initial residual stress state $\underline{\underline{\sigma}}^0$ of the artery and no other boundary condition was applied. During the simulation, the needle-maintained specimen was left free to find its equilibrium $\underline{\underline{\sigma}}^{1\,mod}$, thus simulating the experimental step of cutting it out of the artery. The same procedure as detailed above was used to extract average stress values as a function of r.

The process of identifying the initial residual stress field $\underline{\underline{\sigma}}^0$ was driven by the minimization of the following cost function C which is a measure of the distance between the model and experimental $\underline{\underline{\sigma}}^1$ fields:

$$\min_{\underline{\underline{\sigma}}^0} C\left(\underline{\underline{\sigma}}^0\right) \text{ where } C\left(\underline{\underline{\sigma}}^0\right) = \sum_r \sum_k \left[\sigma_k^{1\,mod}(r) - \sigma_k^{1\,exp}(r)\right]^2 \qquad (2)$$

where $\sigma_k$ represents the $k^{th}$ component of the tensor $\underline{\underline{\sigma}}$, k designates the axial and circumferential components and r ranges over the ten layers considered.

To iteratively update the initial residual stress field $\underline{\underline{\sigma}}^0$, a simplified Newton algorithm was used. The gradient used in the usual Newton algorithm was assumed equal to one, which proved to be sufficient for convergence. At each iteration, $\underline{\underline{\sigma}}^0$ was then updated according to the formula:

$$\underline{\underline{\sigma}}^{0(n+1)}(r) = \underline{\underline{\sigma}}^{0(n)}(r) - \left(\underline{\underline{\sigma}}^{1\,mod(n)}(r) - \underline{\underline{\sigma}}^{1\,exp}(r)\right) \qquad (3)$$

where n is the iteration number.

This identification method was initiated by first using the experimental needle-maintained stress field, $\underline{\underline{\sigma}}^{1(exp)}$, as the initial condition of the FE model. Convergence was considered to be reached when the relative error on $\underline{\underline{\sigma}}^1$ was below a threshold, ε, of 6% for axial and circumferential components and for any radius (excluding the innermost and outermost radii for reasons already mentioned earlier). The initial residual stress field $\underline{\underline{\sigma}}^{0(n)}$ was then considered to be the initial residual stress field in the artery (see schematic of this inverse method in Fig. 6a).

3. **Results**

Prior to developing the results obtained with the above presented method, note, for comparison, that the residual circumferential Green strains calculated with the opening angle method on the inner and outer surfaces of this specimen are $\varepsilon_{\theta i} = -0.052$ and $\varepsilon_{\theta o} = 0.068$, respectively.

### 3.1. Study of the stress release step

The results obtained for the numerical reconstruction of the needle-maintained stress field are presented in this section. They represent the transformation between the stress-free and needle-maintained configurations.

A map of axial and circumferential strain fields through the specimen is shown in Fig. 7. Except the qualitative distribution of circumferential strain showing a typical bending situation, no clear trend could be distinguished from this type of visual result. Therefore, the distribution of strains was also averaged at different radial coordinates and plotted as a function of the radial coordinate (see Fig. 8a), which is more explicit. Remember that edge and corner effects prevent from considering the results close to the inner and outer radius values. Keeping this in mind, this graph emphasizes an almost linear increase of the circumferential strain component with increasing radius, until 75% of the total thickness. Beyond this radius value, it seems to stabilize. The circumferential strain at the inner radius is about $-0.75 \times 10^{-2}$, which corresponds to a compressive strain, and it increases up to values around $4.3 \times 10^{-2}$, corresponding to a tensile strain. Note that the zero circumferential strain is obtained at 25% of the total thickness, not at mid-thickness. The axial strain is positive, ranging from $2.4 \times 10^{-2}$ to $3.2 \times 10^{-2}$, and increases in a linear fashion until 60% of thickness, then it seems to remain almost constant until the outer radius.

The distribution of the computed residual stresses is plotted in Fig. 8b. When assuming the constitutive model of the material to be homogeneous, as was done here, the axial and circumferential residual stress components show similar evolutions. They increase linearly through the thickness until 75% of the total thickness and then they seem to stabilize. This trend is similar to that of axial and circumferential strain evolutions. A note has to be mentioned about the radial response. Radial strains simply accommodate for incompressibility, and it is not judged relevant to comment it. The radial stress component is not commented either as its value is negligible due to free boundary conditions in this direction.

The graph in Fig. 8a raises, however, an important issue. The stress state of the specimen is not at equilibrium; otherwise the integral of the circumferential and axial stresses through the thickness would be zero because the specimen is supposed to be free of any external load. This effect is likely to be related to the application of foam inside the vessel segment during the experimental procedure. Though its ends were open, this segment was fairly long. As a result of the foam expanding both radially and axially, a slight pressure and a slight axial friction-induced stress may have remained in the central cross-section from which the specimen was extracted. For this reason, the curves were shifted towards the left by an amount guaranteeing a zero integral. Here, the offset was respectively 2.1 and 2.39 kPa in the circumferential and axial directions, which may reasonably be attributed to the action of the foam onto the inner surface of the specimen. The corresponding graph is plotted in Fig. 9. The so-obtained stress field, function of r, was considered to be the experimental needle-maintained stress field in the subsequent identification of the residual stress field.

### 3.2. Residual stress field identification

The iterative process presented in section 2.4 successfully converged after 5 iterations. The results of this identification are presented in Fig. 10. This graph emphasizes the needs for the third step of the proposed method consisting in an inverse identification of the initial residual stress field. Indeed, although the trend remains the same regarding the evolution of axial and circumferential stresses through the arterial wall, it is important to notice that the values significantly changed between the needle-maintained and residually-stressed configurations. They are almost doubled at any radius of the arterial wall.

One can observe quasi-linear evolutions for both stress components, with compression stresses below mid-thickness and tension stresses above. This increase seems, however, to be less pronounced beyond 75% thickness. The maximum absolute values reached by each component are similar in tension and compression: about 2 kPa for the axial component, and 3 kPa for the circumferential one.

Note that the results for strain changes between the needle-maintained and residually-stressed configurations (not shown here) were found to be negligible compared to those found between the needle-maintained and stress-free configurations (by two orders of magnitude). A significant amount of the initial stress is transferred to the needles during the cutting-out step, but almost no deformation is observed as the level of these stresses is very low.

## 4. Discussion / conclusion

The present work was intended to address the question of residual stresses in arteries from an original point of view. A new experimental method based on contour methods was setup to investigate the residual strain state of arterial walls at a local scale. A substantial numerical post-processing method was developed to identify the residual stress of the arterial specimen.

The overall method is made up of a first experimental step, performed at the scale of the arterial wall, to obtain an intermediate state marked by the initial residual state. This is followed by a numerical reverse reconstruction of this intermediate stress state and then an inverse reconstruction of the initial stress state. Only a few experiments have been carried out so far and no quantitative conclusion can be drawn for the moment, the experimental approach needing to be further validated. However, the new protocol proposed in this study, based on imaging and finite element analyses, has revealed important subjects of discussion that could be investigated in the future regarding residual stresses in arteries.

The values of circumferential residual strain at the inner and outer radius derived from the optical measurements are in the range of order of typical values reported in the literature for mammal arteries. Note, however, that the raw measured values are found to be slightly lower than the values obtained with the opening angle method. This finding, however, seems not to be due to the cutting-out step which induces negligible strains as mentioned in section 3.2.

More importantly, this study revealed some qualitative trends that should be investigated in deep details. Especially, the distribution of circumferential and axial residual stresses is not necessarily linear through the whole thickness of the arterial wall. It was found that both axial and circumferential residual stresses tend to increase less beyond 75% of the total thickness. This interesting finding could be related to the microstructure of the different layers. Indeed, it is known that the arterial wall is made up of three layers [3]. The inner layer, the intima, has a negligible thickness compared to the other two layers when healthy. The outer layer, the adventitia, consists primarily of a loose network of collagen fibers. The mid layer, the media, contains mainly smooth muscle cells, elastin fibers organized in sheets and some collagen fibers. An interesting histologic illustration of this structure can be found in Fig. 14 in [9].The media/total thickness ratio usually ranges from 0.5 to 0.9. Though this ratio was not precisely measured in this experiment, our results show a clear change in the evolution of residual strain and stress from 75% of the total thickness. From this radial location and up to the external surface, both axial and circumferential strains and stresses seem to stabilize or increase less. A possible explanation for this would be that the residual mechanical state of the arterial wall is mainly governed by the medial layer and it is propagated in the adventitial layer because they are bond together. Such hypothesis is in agreement with previous observations by Holzapfel et. al. [9] which suggested that the medial layer is dominant in residual stress-induced bending. This phenomenon may be related to the histological content of the arterial wall as it is known that elastin and smooth muscle cells provide most of the structural stiffness of the media, whereas the loose collagen fibers (see Fig. 14 in [9]) in the adventitia provide stiffness only at large stretch ratios [24].

Another important finding of this study is that the residual stress field is confirmed to be two-dimensional, and not only circumferential. This consideration is generally not taken into account in the literature and may affect the distribution of residual stresses through the thickness of arteries due to Poisson's effects. This fact was earlier evidenced in [9] but it was never measured simultaneously on a single specimen. Experimental methods using arterial strips with one direction of the specimen being large compared to the other are indeed not appropriate to measure strains in both directions with comparable resolution and accuracy. Contrarily to other methods, the strength of the present approach is that it was conducted on a specimen with comparable circumferential and axial dimensions. This is favorable to the measurement of both the circumferential and axial residual strains. We showed thereby that axial residual strain is not negligible and moreover, it affects the axial residual stress which was shown to increase linearly until about 75% of the total thickness.

Another aspect that may affect the residual stress state is the constitutive behavior which may be different in medial and adventitial layers, even at low strain. The rationale supporting this assumption relies on the nature of the composition of these layers that was already mentioned above. The adventitia comprises mainly collagen fibers which are loose at the no-pressure state, whereas the media includes elastin sheets, smooth muscle cells and other components which provide an elastic stiffness even at low pressures. For this reason, different constitutive parameters may be allocated to each of these layers, the adventitia being softer [10]. This may increase the complexity of the residual-stress distribution in the tissue, but it is likely to reinforce the hypotheses made above.

Investigating the latter point more thoroughly will first require the development of a new theoretical background which is out of scope of this paper. A recent work from Holzapfel [10] constitutes a first attempt for addressing this issue. Confirming the level of axial residual stress will also require applying the imaging and inverse finite element approach developed in this study on a large number of specimens to generate a sufficient database. To this aim, a certain number of aspects regarding tissue handling will be improved for rendering the approach more robust. In particular the foam influence that may induce some internal pressure and an offset on the residual stresses will be investigated. Regarding the finite element model, it is envisaged including more realistic, and possibly layer-specific, constitutive equations based on local indentation tests for deriving the values of residual stresses in future experiments.

5.  **Conclusion**

A novel methodology to measure the residual strain state of arteries at the scale of its wall has been developed and presented throughout this paper. The method consists in extracting a cube-shaped specimen out of the arterial wall and performing whole-body imaging of it while maintaining it in a temporary intermediate stress state. Then the specimen is left free so as to release internal stresses, and imaged again. Using DIC processing, the surface displacement field is reconstructed during stress release and further used in FE models to compute the whole residual stress field of the specimen existing before extraction out of the artery.

The method evidenced the three-dimensional nature of the residual state of load-free arteries. Axial and circumferential residual strains and stresses do not necessarily vary linearly across the whole thickness of the specimen, which might be closely related to the inhomogeneous nature of the arterial wall. Though improvements and validation of the technique are still required, a proof of feasibility was shown, revealing intriguing problems posed by the local distribution of residual stresses in arteries.

**6.     Figure captions**

Fig. 1: Global outline of the residual stress identification method developed in this study. The top line represents the experimental steps of the method while the bottom line represents the numerical steps.

Fig. 2: Sample preparation: a) the artery was filled with expandable sealant foam and the foam was allowed to harden; b) the areas of interest were delimited by using acupuncture needles and the remaining part of the artery was cut and then removed; c) the foam was cut into the final shape.

Fig. 3: Scheme of the DIC-based whole-body measurement set-up used for the experiments. The gooseneck-pipe illumination system has not been reported for clarity of representation.

Fig. 4: Two images of the same face of the sample before (a) and after releasing residual stresses (b). Plot of the corresponding surface y-displacement map showing the typical distribution of a beam in pure bending (c).

Fig. 5: Data points obtained from the 'whole-body' DIC contouring of the stress-free configuration. The use of a single 'cylindrical' dot calibration pattern glued on the post and always present in the image (see Fig. 3 a, b) allowed to automatically reconstruct and merge the data points of the 8 stereo-systems in a single reference system. The corresponding idealized FE model geometry is shown in (b).

Fig. 6: Inverse identification of the residual stress field. (a) Schematic of the method. $\underline{\underline{\sigma}}$ denotes a symmetric second-order tensor composed of six components $\sigma_k$ which are function of the radial component r. (b) Transparent, schematic, view of the FE model of the cutting-out procedure including eight needles through the arterial continuum.

Fig. 7: Logarithmic strain fields, in the needle-maintained configuration, shown in a cross-sectional view of the model: circumferential strain (a) and axial strain (b).

Fig. 8: Logarithmic strain (a) and Cauchy stress (b) distribution along the radial coordinate, in the needle-maintained configuration.

Fig. 9: Cauchy stress distribution along the radial coordinate, in the needle-maintained configuration, with respective corrective shifts of -2.1 and -2.39 kPa assumed to correspond to the action of the foam in the circumferential and axial directions.

Fig. 10: Residual Cauchy stress distribution obtained with the inverse identification approach, compared to the needle-maintained stress distribution.